\documentstyle[12pt]{article}
\begin{document}
\baselineskip=1.2\baselineskip

\def\bea{\begin{eqnarray}}
\def\eea{\end{eqnarray}}

\renewcommand{\thefootnote}{\fnsymbol{footnote}}
\setcounter{equation}{0}
\setcounter{section}{0}
\renewcommand{\thesection}{\arabic{section}}
\renewcommand{\theequation}{\thesection.\arabic{equation}}

\pagestyle{plain}
\begin{titlepage}
\hfill{NUS/HEP/95-11}
\begin{flushright}
hep-th/yymmnn
\end{flushright}
\begin{center}
{\large{\bf { Asymptotically Free $\hat{U}(1)$ Kac-Moody Gauge Fields
in $3+1$ dimensions. }}}\par
\end{center}
\vskip 1.2cm
\begin{center}
{ Belal E. Baaquie\footnote{E-mail: phybeb@nus.sg}
and Rajesh R. Parwani\footnote{On leave from
Institute of Physics, Bhubaneswar, India.  E-mail: parwani@iopb.ernet.in} \\
Department of Physics, National University of Singapore,\\
SINGAPORE 0511}
\end{center}
\vskip 1.0cm
\centerline{ November 1995}
\vskip 1.0cm
\centerline{PACS 11.10.Lm, 11.15.-q, 11.30.Ly, 11.10.Jj,
11.10.Gh.}
\vskip 2.5 cm
\centerline{\bf Abstract}
{$ \hat {U}(1)$ Kac-Moody gauge fields have the infinite dimensional
$ \hat{U}(1)$
Kac-Moody group as their gauge group.  The pure gauge sector, unlike the
usual $U(1)$ Maxwell lagrangian, is nonlinear and nonlocal; the
Euclidean theory is
defined on a $d+1$-dimensional manifold
$  {\cal{R}}_d \times {\cal{S}}^1  $
and hence is also asymmetric.  We
quantize this theory using the background field method and  examine its
renormalizability at one-loop by  analyzing all the relevant diagrams.
We find that, for a suitable choice of the gauge field propagators,
this theory is one-loop renormalizable in $3+1$ dimensions.  This
pure abelian Kac-Moody gauge theory
in $3+1$ dimensions has only one running coupling constant
and the theory  is asymptotically free. When fermions are added the
number of independent couplings increases and a richer
structure is obtained. Finally, we  note some features of the
theory which suggest its possible relevance to the study of
anisotropic  condensed matter systems, in particular that of
high-temperature superconductors.}
\end{titlepage}

\newpage

\section{Introduction}

Kac-Moody groups embody the infinite dimensional symmetries that
underlie string theory and conformal field theory \cite{PG}.  It  is
natural to realize the Kac-Moody group $\hat{G}$ as a gauge
symmetry, that is as a  group of gauge
transformations of gauge fields, and to obtain a lagrangian invariant
under these gauge transformations.

A lagrangian of vector gauge fields coupled to matter fields that is
invariant under Kac-Moody gauge transformations for an arbitrary group
$\hat{G}$ was obtained in \cite{B1}.  A Kac-Moody group\footnote{The
"Kac-Moody group" we refer to here is sometimes called
 an "affine Lie Group", and the corresponding algebra is known
to physicists as current algebra.}
 $\hat{G}$ consists
of all mappings of the
circle ${\cal{S}}^1$  into the compact Lie group $G$.
Let ${\cal{R}}_d$ be an Euclidean space consisting of one time
dimension and $(d-1)$ space dimensions.
The
gauge transformation at a point $ x \in {\cal{R}}_{d} $, namely $ \Phi(x) $,
is an element of $\hat{G}$ and has the form ($\int \equiv
\int_{0}^{R} d \sigma $, that is, ${\cal{S}}^{1}$ has radius $ R /2\pi$)
\bea
{\Phi(x)} &=& e^{i \Lambda(x)} \
e^{i \int \phi^{\alpha}(x,\sigma) \ Q^{\alpha}(\sigma)} \,\, .
\label{gt}
\eea
${Q^{\alpha}(\sigma)}$ are the generators of  $\hat{G}$,
$\phi^{\alpha}(x,\sigma)$ and $\Lambda(x)$ are dimensionless
gauge functions,
and $\alpha$ is the Lie group index.
We can consider $\sigma \in {\cal{S}}^1$ to be a new space-like
direction in the theory, which is then effectively defined on
 ${\cal{R}}_d \times {\cal{S}}^1$.  The ${\cal{S}}^1$ direction
has a special significance since each point in  ${\cal{S}}^1$
carries its own non-commuting charge $Q(\sigma)$.
Hence non-abelian gauge fields $A_{\mu}^{\alpha}(x,\sigma)$
are defined on the manifold  $  {\cal{R}}_d \times {\cal{S}}^1  $.
Since
$\Phi(x)$ has a central extension, there is also  a $U(1)$ gauge field
$B_{\mu}(x)$ defined on only $ {\cal{R}}_d.$

Just as for the non-abelian case,
the lagrangian for  $ \hat{U}(1) $ Kac-Moody gauge fields is
nonlinear, nonlocal and asymmetric.  Quantizing the abelian
 theory  is of
particular importance because: (i) this is the starting point for the
analysis of the more complex non-abelian Kac-Moody gauge fields
\cite{B1} and,
(ii) the new nonlinearities arising from the Kac-Moody nature of the
gauge group can be studied in their simplest and most
essential form for the  $ \hat{U}(1) $ case.

In Section II we review the $ \hat{U}(1)$ Kac-Moody gauge symmetry
 and derive the lagrangian; in Section III we quantize the
theory via the Feynman path integral using the background field method.
 In Section IV we compute the one-loop
$\beta$-function, and in Sect.V we
analyze all the one loop Feynman diagrams, identify
the divergent diagrams and show
that the theory is one-loop renormalizable in 3+1 dimensions.  In
Section V we draw some conclusions and highlight a possible application
of the theory as an effective low-energy description of anisotropic
superconductors.

\setcounter{equation}{0}
\section{The Kac-Moody Gauge-Field Lagrangian}

The $\hat{U}(1)$ Kac-Moody generators satisfy the commutation relation
\bea
[Q(\sigma), Q(\sigma^{\prime})] &=& i \kappa \delta^{\prime}(\sigma
- \sigma^{\prime})
\label{al}
\eea
where the prime symbol ($\prime$) on functions means $\partial /
\partial \sigma$ , and $\kappa$ is a real number.
For gauge transformations given by (\ref{gt}), we have
\bea
\Phi(x) \ Q(\sigma) \ \Phi^{\dag}(x) &=& Q(\sigma) + \kappa \phi^{\prime}
(x,\sigma) \label{gt2} \, .
\eea
To determine the  gauge transformation of the gauge-fields,
consider the link variable ($a$=
lattice spacing)
\bea
U_{\mu}(x) = e^{ia g B_{\mu}(x) + ia e \int A_{\mu}(x, \sigma) Q(\sigma)}
\eea
where $g$ and $e$ are dimensionful coupling constants.
We define Kac-Moody gauge transformations by
\bea
U_{\mu}(x) &\to& \Phi(x) \ U_{\mu}(x) \ \Phi^{\dag}(x+a \hat{\mu})
\label{kmgt}
\eea
and in the $a \to 0$ limit we obtain
\bea
\delta A_\mu(x, \sigma) &=& - {1 \over e} \partial_{\mu} \phi(x,\sigma)
\; , \nonumber \\
&& \label{dab}\\
\delta B_\mu(x) &=&  - {1 \over g}
\partial_{\mu} \Lambda(x) + { \kappa e \over g } \int
\phi^{\prime}(x, \sigma) A_\mu(x, \sigma) -
{\kappa \over 2g} \int \phi^{\prime}(x, \sigma)
\partial_{\mu} \phi(x, \sigma) \,\, . \nonumber
\eea
Consider next the open plaquette starting and ending at $x$, that is,
\bea
W_{\mu \nu}(x) &\equiv& U_{\mu}(x) U_{\nu}(x+a \hat{\mu}) U^{\dag}_{\mu}(x+a
\hat{\nu}) U^{\dag}_{\nu}(x) \; .
\eea
To leading order in $a$ we have
\bea
W_{\mu \nu}(x) &=& e^{ia^2 g \chi_{\mu \nu}} e^{ia^2 e \int F_{\mu \nu}
(x, \sigma) Q(\sigma)}  \; ,
\eea
where the pure phase is
\bea
\chi_{\mu \nu} &=& f_{\mu \nu} + {\kappa e^2 \over g}
\int A_{\mu}^{\prime} A_{\nu}
\eea
with
\bea
f_{\mu \nu} &=& \partial_{\mu} B_{\nu} - \partial_{\nu} B_{\mu} \,\, ,
 \nonumber
\eea
and
\bea
F_{\mu \nu} &=& \partial_{\mu} A_{\nu} - \partial_{\nu} A_{\mu} \,\, .
\nonumber
\eea
Note that the term $\kappa \int A_{\mu}^{\prime} A_{\nu}$ in the phase
$\chi_{\mu  \nu} $
arises due to the central extension
in the Kac-Moody algebra and is absent for gauge theories based on
compact Lie groups.
Under gauge-transformations (\ref{kmgt}) we have, upon using
(\ref{gt2}),
\bea
W_{\mu \nu}(x) &\to& \Phi(x) W_{\mu \nu}(x) \Phi^{\dag}(x) \label{wgt}\\
&& \nonumber \\
&=& e^{i a^2 g \chi_{\mu \nu}} e^{i a^2 e ( \int F_{\mu \nu} Q + \kappa
\int F_{\mu \nu} \phi^{\prime})}
\eea
which yields
\bea
\delta F_{\mu \nu} &=& 0 \label{df}
\eea
and
\bea
\delta \chi_{\mu \nu} &=& { \kappa e \over g}
\int F_{\mu \nu} \phi^{\prime} \,\, .
\label{dchi}
\eea
Note that unlike the case of compact Lie groups, the Kac-Moody
gauge transformation (\ref{wgt}) induces
an inhomogeneous transformation of
$\int F_{\mu \nu} \phi^{\prime}$ on $F_{\mu \nu}$ and hence induces
a change in the phase $\chi_{\mu \nu}$.
To obtain a gauge-invariant lagrangian we need to introduce a new field
$A_{5}(x, \sigma)$ to cancel the variation $ \int
F_{\mu \nu} \phi^{\prime}$,
and which transforms as
\bea
\delta A_{5} (x, \sigma) &=& - {1 \over e} \phi^{\prime}(x, \sigma) \, .
\label{da5}
\eea
Defining the antisymmetric tensor
\bea
\Gamma_{\mu \nu}(x) = \chi_{\mu \nu}(x) + { \kappa e^2 \over g}
\int F_{\mu \nu}
A_{5}(x,\sigma) \; ,
\eea
we have from (\ref{df}, \ref{dchi}), and (\ref{da5})
\bea
\delta \Gamma_{\mu \nu} &=& 0 \; .
\eea
It has been shown in \cite{B1} that $\Gamma_{\mu \nu}$ can be
generalised to the nonabelian case.
Define
\bea
F_{\mu 5}(x, \sigma)= \partial_{\mu} A_{5} - A_{\mu}^{\prime}
\eea
which from (\ref{dab}) and (\ref{da5}) is gauge invariant,
\bea
\delta F_{\mu 5} &=& 0 \,\, .
\eea
Hence we have the following classical lagrangian\footnote{Following
the convention of \cite{B1}, the index $\mu$ covers the $d$ coordinates
of the Euclidean manifold
${\cal{R}}_d$ while $'5'$ labels the $(d+1)$-th
coordinate in ${\cal{S}}^1$.}
in ${\cal{R}}_{d} \times {\cal{S}}^{1}$ which is invariant
under the Kac-Moody gauge transformations (\ref{dab}) and (\ref{da5})
\bea
{\cal{L}}(x) &=& {1 \over 4} \Gamma_{\mu \nu}^{2}(x)
+ {1 \over 4} \int d\sigma d\sigma^{\prime} \ F_{\mu \nu}(x, \sigma)
h(\sigma-\sigma^{\prime}) F_{\mu \nu}(x, \sigma^{\prime}) \nonumber \\
&& \label{class} \\
&& \; +
{1 \over 2} \int d\sigma d\sigma^{\prime} \ F_{\mu 5}(x, \sigma)
f(\sigma-\sigma^{\prime}) F_{\mu 5}(x, \sigma^{\prime}) \,\, ,
\nonumber \eea
and where, as defined above ($\partial_5 = \partial / \partial \sigma$),
\bea
\Gamma_{\mu \nu} &=& \partial_{\mu}B_{\nu} - \partial_{\nu}B_{\mu}
+ {\kappa  e^2 \over g}
\int  \ A_{\mu}^{\prime} A_{\nu} \ + \ { \kappa e^2 \over g}
\int  F_{\mu \nu} A_{5} \, , \nonumber \\
F_{\mu \nu} &=& \partial_{\mu} A_{\nu} - \partial_{\nu} A_{\mu} \, ,
 \label{ccgt} \\
F_{\mu 5} &=& \partial_{\mu} A_{5} - \partial_{5} A_{\mu} \, \nonumber.
\eea
Recall that $A_{\mu} = A_{\mu}(x, \sigma), \ A_{5}= A_{5}(x, \sigma)$
and $B_{\mu} = B_{\mu}(x) $. Since $\Gamma_{\mu \nu}(x)$ has an
asymmetric coupling in the $\sigma$-direction, we anticipate a similar
asymmetry in the pure $A_\mu , A_5$ sector and this is reflected in the
functions $h$ and $f$ which have been introduced in (\ref{class}).
These functions are required to be positive but are otherwise
undetermined.

{}From the kinetic term for the $B_{\mu}$ field, its canonical mass dimension
is deduced to be $(d-2)/2$. We choose the $A_{\mu}$ and $A_{5}$ fields
to have the mass dimension $(d-1)/2$, which is the appropriate canonical
dimension for gauge fields in $d+1$ dimensions.
Then the coupling constant $e$ has mass dimension $(3-d)/2$ and
\bea
\lambda &\equiv& {\kappa e^2 \over g}
\eea
has mass dimension $(2-d)/2$, while the mass dimension of the
functions $h(\sigma)$ and $f(\sigma)$
in (\ref{class})
is $1$. The Fourier transform of $h(\sigma)$
(and similarly for $f(\sigma))$ can therefore be written as
($\sum_{n} \equiv \sum_{n= - \infty}^{\infty}$)
\bea
h(\sigma) &=& { 1 \over R} \sum_{n} \
h_{n} \ e^{i p_{n} \sigma} \; ,
\eea
where $p_n = 2 \pi n /R , \ n \in {\cal{Z}},$ $R$ is the length
of the compact $\sigma$-dimension, and the Fourier coefficients
$h_n$ are dimensionless.
We will eventually be interested in the case $d=3$ and since
$R$ is the natural scale in the theory,  we can define
a
dimensionless coupling constant
\bea
\bar{\lambda} &=& {\lambda}/ \sqrt{R}.
\eea

Though we will mostly discuss the pure gauge theory in this paper,
we summarize  here the coupling of fermions to the gauge fields
\cite{B1}. Their gauge-invariant coupling to the fields $A_\mu, A_5$
is given
by the usual Dirac lagrangian\footnote{Note again that the index $'5'$
refers to the compact coordinate, and so $\gamma_5$ has nothing to do
with the usual chirality matrix.}
\bea
S_{1}^{F} &=& \int d^d x \ d \sigma \ \bar{\psi}(x, \sigma) \ [ (
\partial_{\mu} -ie A_{\mu}) \gamma_{\mu} \ + \
(\partial_5 -ie A_5)\gamma_5 \ + \  m ] \ \psi(x, \sigma)
\label{1f}
\eea
where $\psi(x, \sigma),\bar{\psi}(x, \sigma) $ are the Dirac fermions
defined on the $(d+1)$ dimensional manifold
${\cal{R}}_d \times {\cal{S}}^1$.
The coupling of fermions to $B_\mu(x)$
is more subtle \cite{B1}. Consider the $\hat{U}(1)$ super
Kac-Moody algebra in which Majorana fermions $H(\sigma)$, satisfying
$\{ H(\sigma), H(\sigma^{\prime}) \} = \delta(\sigma
-\sigma^{\prime})$, together with $Q(\sigma)$ form the generators.
The super Virasoro algebra formed from
$Q(\sigma)$ and $H(\sigma)$ has $c=1+1/2$ with a (suitably regularized)
super-charge operator $G(\sigma) = {1 \over \kappa} Q(\sigma) H(\sigma)$.
Kac-Moody fermions $\Psi(x)$ are defined on
${\cal{R}}_d$, and form an arbitrary irreducible representation of the super
Virasoro algebra. They have Kac-Moody invariant coupling to the
gauge fields as follows :
 \bea
S_{2}^{F} &=& \int d^d x  \ \bar{\Psi}(x) \ [ (
\partial_{\mu} -ie \int A_{\mu}(x,\sigma) Q(\sigma) -igB_\mu(x))
\gamma_{\mu}
\ + \ M ] \ \Psi(x) \nonumber \\
&&  \label{2f} \\
&& \,\,\, + \ \tau \int d^d x \ \Psi(x) \ \int [ G(\sigma) - A_5(x,\sigma)
H(\sigma) ] \ \Psi(x) \,\, . \nonumber
\eea
The $\tau$ term in $S_{2}^{F}$ has been introduced to attenuate the
high mass states.
The full fermion coupling is given by $S_{1}^F + S_{2}^F$.
The thermodynamical partition function  of
the free lagrangian for $\Psi(x)$ given in (\ref{2f}) has been studied in
\cite{B2} and exhibits a maximum limiting temperature.
In Section 4.4 we will discuss the contribution of the ordinary
Dirac fermions
(\ref{1f}) to the beta-function.

\setcounter{equation}{0}
\section{Path-Integral Quantization.}

We define the quantum theory through the Feynman path-integral and the
Fadeev-Popov { \it ansatz} for the gauge-fixing.
It is convenient to use the background field formalism \cite{HA}
whereby the
fields are split into background fields ($A_{\mu}, A_5, B_\mu$)
and quantum fields ($a_\mu, a_5, b_\mu$). The classical action
(\ref{class}) is then
invariant under the transformation
\bea
\delta(A_5 + a_5) &=& - {1 \over e} \phi^{\prime} \, , \label{split1}\\
\delta(A_\mu + a_\mu) &=& - {1 \over e} \partial_{\mu} \phi \, , \\
\delta(B_\mu + b_\mu) &=&  -{1 \over g}
\partial_{\mu} \Lambda + {\kappa e \over g} \int
\phi^{\prime} (A_\mu + a_\mu) - {\kappa  \over 2g} \int \phi^{\prime}
(\partial_{\mu} \phi) \,\, . \label{split}
\eea
The action with  given background fields is still invariant
under a gauge-transformation of the quantum fields alone.
These quantum gauge transformations are deduced by
deleting the classical
fields on the left-hand-side of (\ref{split1}-\ref{split}),
\bea
\delta^{Q} a_5 &=& - {1 \over e} \phi^{\prime}  \, , \\
\delta^{Q} a_\mu &=& - { 1 \over e} \partial_{\mu} \phi \, , \\
\delta^{Q} b_\mu &=& -{1 \over g}  \partial_{\mu} \Lambda +
{ \kappa e \over g} \int
\phi^{\prime} (A_\mu + a_\mu) - {\kappa  \over 2g} \int \phi^{\prime}
(\partial_{\mu} \phi) \,\, . \label{qsplit}
\eea
Hence to define
the background field path-integral $Z[A_\mu , A_5, B_\mu]$,
the quantum gauge fields must be gauge-fixed and an infinite
group volume factored out. However
the gauge-fixed action
is required to be invariant under
background field transformations which are of the same form as the
classical  gauge-invariance. That is,
\bea
\delta^{B} A_5  &=& - {1 \over e} \phi^{\prime} \; ,\label{bs1} \\
\delta^{B} A_\mu &=& - {1 \over e} \partial_{\mu} \phi \; , \label{bs2} \\
\delta^{B} B_\mu  &=& -{1\over g}  \partial_{\mu} \Lambda +
{ \kappa e \over g} \int
\phi^{\prime} (A_\mu ) - {\kappa  \over 2g} \int \phi^{\prime}
(\partial_{\mu} \phi) \; , \label{bs3} \\
\delta^{B} b_\mu &=& {\kappa e \over g} \int a_\mu \phi^{\prime}
\label{bs4}\\
\delta^{B} a_5 &=& \delta^{B} a_{\mu} = 0 \,\,\, . \label{bs5}
\eea
A gauge-fixing term for the $a_{\mu}$ and $b_{\mu}$ fields
  which is invariant under (\ref{bs1}-\ref{bs5}) is
\bea
 \delta(\partial_\mu a_\mu) \ \delta \left(\partial_\mu b_\mu + {\kappa e^2
\over g} \int
A_{\mu}^{\prime} a_\mu - c_{\mu} \right)
\label{gf}
\eea
with $c_{\mu}(x)$ an arbitrary function. To see the invariance of
the Dirac-delta-function constraints in (\ref{gf}),
consider the variation of the argument of the second term,
\bea
\delta^{B} \left( \partial_\mu b_\mu + {\kappa e^2 \over g} \int
A_{\mu}^{\prime} a_\mu - c_{\mu} \right)
&=& \partial_\mu \delta^{B} b_\mu + {\kappa e^2 \over g} \int
( a_\mu \delta^{B} A_{\mu}^{\prime} + A_{\mu}^{\prime}
\delta^{B} a_{\mu}) \nonumber \\
&=& \partial_\mu \left( {\kappa e \over g} \int a_\mu \phi^{\prime} \right)
 + {\kappa e^2 \over g} \int {(- \partial_{\mu} \phi^{\prime}) \over e}
a_\mu \nonumber \\
&=&  {\kappa e \over g} \int \phi^{\prime} \partial_{\mu} a_{\mu}
\nonumber \\
&=& 0 \,\, , \nonumber
\eea
where in obtaining the last line we used that, from (\ref{gf}),
$\partial_{\mu} a_{\mu} =0$. Also, the constraint on the $a_\mu$
field in (\ref{gf}) is trivially invariant under  (\ref{bs5}).
When the gauge-fixing (\ref{gf}) is implemented in the path-integral, the
constraint on the $b_{\mu}$ field may be exponentiated as usual by
averaging over the $c_{\mu}$ variable
with a Gaussian weight. However, as noted above, it is crucial for the
invariance of (\ref{gf}) that the $a_{\mu}$ field be in the
strict Landau gauge. The Fadeev-Popov ghosts due to the above gauge-fixing
are non-interacting and so may be ignored.
Thus the gauge-fixed action, including external sources $J_{\mu}, K_{\mu}$
and $L$, is
\bea
S_{eff} &=&
\lim_{\alpha \to 0} \int d^{d} x \ \left\{ {\cal{L}}(A_5 + a_5, A_{\mu}
+ a_{\mu}, B_{\mu} + b_{\mu}) + { 1 \over 2 \alpha} \int (\partial_{\mu}
a_{\mu})^{2} \right\} \nonumber \\
&& \\ \label{qact}
&& \;\; +  \int d^d x \ \left\{ {1 \over 2 \eta} ( \partial_{\mu} b_{\mu}
+ {\kappa e^2 \over g} \int A_{\mu}^{\prime} a_{\mu})^{2}
\ + \ \int J_{\mu} a_{\mu} \ + \ \int L a_5 \ + \ K_{\mu} b_{\mu}
\right\} \,\, , \nonumber
\eea
with the lagrangian ${\cal{L}}$ given by (\ref{class}-\ref{ccgt}).
This action is invariant under the background transformation (\ref{bs1}
-\ref{bs5})
plus the following transformation for the sources
\bea
\delta^{B} J_{\mu} &=& -{ \kappa e \over g} K_{\mu} \phi^{\prime}
\, ,  \nonumber \\
&& \label{bs6} \\
\delta^{B} K_{\mu} &=&  \delta^{B} L =0 \,\, ,
\nonumber
\eea
and therefore the quantum field theory defined by the background
path-integral
\bea
Z[A_\mu, A_5, B_\mu, J_\mu, K, L_\mu ]&=& \int [Da_\mu \ Da_5 \ Db_\mu]
\ e^{-S_{eff}}
\label{into}
\eea
is invariant under the transformations  (\ref{bs1}-\ref{bs5})
and (\ref{bs6}).
The one-particle-irreducible (1PI) effective action of the "vacuum" theory
(that is without background fields)
is obtained
by computing, using (\ref{into}), 1PI graphs with no external
quantum fields \cite{HA} and by identifying the background fields in
(\ref{qact}) with the  fields appearing in the 1PI action.
 This 1PI action is invariant under
the Kac-Moody gauge transformations (\ref{bs1}-{\ref{bs3}).

For a one-loop calculation we only need to expand
(\ref{qact}) up to second order in the quantum fields. Terms linear in
the quantum fields may be ignored as they do not contribute to the
1PI effective action.
The free propagators for the gauge-fields are given in momentum space by
(for simplicity we choose the Feynman gauge $\eta=1$ for the
$b_{\mu}$ field)
\bea
<b_{\mu} b_{\nu}> &=& { \delta_{\mu \nu} \over p^2} \; , \\
<a_{\mu} a_{\nu}> &=& \left(\delta_{\mu \nu} -
{ p_{\mu} p_{\nu} \over p^2}\right) { 1 \over f_n p_{n}^2 + h_{n} p^2}
\; ,  \label{aprop} \\
<a_{5} a_{5} > &=& {1 \over f_n p^2} \; , \\
<a_{\mu} a_{5} > &=& 0 \; .
\eea
Note that because of the strict Landau gauge condition
on the $a_{\mu}$  field, there is no mixed $<a_{\mu} a_5>$
propagator at tree-level.

Finally, the fermionic action (\ref{1f}), being conventional,
can be incorporated in the usual way \cite{HA}.

\setcounter{equation}{0}
\section{Beta function}
\subsection{General}
The tree-level action (\ref{qact})
has been written in terms of "bare"
quantum and background fields, and bare coupling
constants $\lambda$ and $e$.  When the quantum fields
are integrated out in the path-integral (\ref{into})
 to generate the full 1PI effective action,
UV divergences will be present  in the new terms which have been
added
to the classical action.
As the full 1PI action is guaranteed to be
gauge-invariant in the background gauge, the divergences will take
a gauge-invariant form. If the theory is renormalizable, which we show
below to be the case at one-loop, the divergences can be removed
by rescaling the fields and couplings in the bare classical
lagrangian so that
the new lagrangian, and hence the physical partition function,
 is finite when written in terms of finite "renormalized"
quantities. Consider the one-loop UV divergent
correction $\Delta \Gamma_{\mu \nu}^{2}$  to $\Gamma_{\mu \nu}^2$.
Then we want to write
\bea
\Gamma_{\mu \nu}^{2} + \Delta \Gamma_{\mu \nu}^{2} &\equiv& \Gamma_{\mu
\nu}^{(r) 2} \, ,
\eea
by renormalizing the bare tensor
( in $d=3-2\epsilon$ dimensions)
\bea
\Gamma_{\mu \nu} &=& \partial_{\mu}B_{\nu} -
\partial_{\nu}B_{\mu}
+  \mu^{\epsilon}_{0} \ \sqrt{R} \ \bar{\lambda}
 \int d \sigma (\ A_{\mu}^{ \prime} A_{\nu} \ + F_{\mu \nu}
 A_{5}) \, , \label{bare}
\eea
where $\mu_{0}$ is some fixed mass scale (such as $R$),
into the renormalized tensor
\bea
\Gamma_{\mu \nu}^{r} &=& \partial_{\mu}B_{\nu}^{r} -
\partial_{\nu}B_{\mu}^{r}
+  \mu^{\epsilon} \ \sqrt{R} \ \bar{\lambda}_{r}
 \int d \sigma (\ A_{\mu}^{r  \prime} A_{\nu}^{r} \ + F_{\mu \nu}^{r}
 A_{5}^{r}) \label{renorm} \;
\eea
defined at the new mass scale $\mu$.
The bare quantities in (\ref{bare}) are
related to the renormalized ones in (\ref{renorm}) by
the $Z$ factors,
\bea
B_{\mu} &=& \sqrt{Z_{B}} B_{\mu}^{r} \; , \\
A_{\mu} &=& \sqrt{Z_{A}} A_{\mu}^{r} \; , \\
A_{5} &=& \sqrt{Z_{5}} A_{5}^{r} \; ,
\eea
and
\bea
{\bar \lambda} = Z_{\lambda} \bar{\lambda}_{r} \,\,\, .
\eea
(We are implicitly assuming the theory is multiplicatively
renormalizable, a fact which we will verify to one-loop by explicit
calculations.)
As the Kac-Moody gauge-invariance (\ref{bs1}-\ref{bs3}) of the 1PI action
is maintained
in the background gauge, the
$Z$ factors must be related as follows
\bea
Z_5 &=& Z_A \label{za5}
\eea
and
\bea
\sqrt{Z_{B}} &=& Z_{\lambda} \mu_{0}^{\epsilon} \mu^{-\epsilon}
\sqrt{Z_{A} Z_{5}} \\
&=& Z_{\lambda} Z_{5} \left( {\mu_{0} \over \mu} \right)^{\epsilon} \; .
\eea
Hence
\bea
\bar{\lambda} =  {\sqrt{Z_{B}} \over Z_{5}} \left( {\mu \over \mu_{0}}
\right)^{\epsilon}
\ \bar{\lambda}_{r} \label{lo} \, ,
\eea
from which the $\beta$-function $\beta_{\lambda} = \mu { \partial
 \bar{\lambda}_{r} \over \partial \mu}$ be determined by differentiating
both sides of (\ref{lo})
with respect to $\mu$ and taking $\epsilon \to 0.$
In subsections 4.2 and 4.3, we
determine the  wavefunction renormalization factors
required in (\ref{lo}) to one-loop order in the pure gauge theory.
Note that
though $Z_5$ and $Z_A$ are equal from (\ref{za5}), we have used the
former in (\ref{lo})
as it is easier to determine directly (only two diagrams need be analysed).
In Section 4.4 we will discuss how including the Dirac fermions (\ref{1f})
changes the number of independent couplings, and also
compute the new one-loop beta-functions.\\

\subsection{$Z_B$}
Two diagrams (see Fig.2)
contribute to the one-loop correction $\Delta \Gamma_{\mu \nu}^{2}
$ and these determine $Z_B$.
We obtain
\bea
\Delta \Gamma_{\mu \nu}^{2}(x) &=& -4 {\cal{T}}
\Gamma_{\mu \nu}^2(x)
\eea
where
\bea
\cal{T}&=& {-\bar{\lambda}^{2} R \over 8} (4-d) \ {\cal{S}}(d) \
\int {d^d q \over  (2\pi)^d}
 {1 \over q^2 (q^2 +1)} \ + \ \mbox{convergent terms as } \ d \to 3 \, ,
\nonumber
\eea
\bea
\mbox{and } \;\;\;\;\; {\cal{S}}(d) &\equiv& \mu_{o}^{2 \epsilon}
\left({2\pi \over R}\right)^{d-2} \sum_{n}
{|n|^{d-2} \over h_{n}^{d/2}} f_{n}^{d-4 \over 2} \;\; .
\label{sum}
\eea
Note that
although the functions $f_n$ and $h_n$ appear in very different ways in
the two diagrams, they  finally contribute in the
same manner through the sum above.
Since the $d$-dimensional loop momentum integral is ultraviolet (UV)
finite for $d \le 3$,
the only possible ultraviolet (UV)
divergence as $d \to 3$ comes from the sum over $n$.
To evaluate $(\ref{sum})$
we have to specify the functions $f_n$ and $h_n$. We choose
(see below)
\bea
h_n &=& c_{1} |n| + c_{1}^{\prime} \, , \label{c1} \\
f_n &=& c_{2} |n| + c_{2}^{\prime} \, \label{c2} .
\eea
For the boundedness  of the path-integral we require the constants
$c_i$ and $c_{i}^{\prime}$ to be non-negative.
If either of $c_1$ or $c_2$ were zero then, since
the Riemann zeta-function $\zeta(s) = \sum_{1}^{\infty} {1 \over n^s}
$ has a pole only for $s=1$, (specifically $\zeta(s \to 1) \to 1/(s-1))$,
the sum (\ref{sum}) would be  finite as $d \to 3$.
For the choice $(\ref{c1}-\ref{c2})$ with $c_1$ and $c_2$ nonzero,
and as $\epsilon \to 0$,
\bea
S(d=3-2\epsilon) &\to& {2 \pi \over R}
{1 \over \sqrt{c_{1}^{3} c_2 } } \ { 1 \over \epsilon} +
O(\epsilon^{0})
\,\, .
\eea
Hence, using the minimal subtraction scheme,
\bea
\Delta \Gamma_{\mu \nu}^{2}(x) &=&  {\bar{\lambda}^{2}
\over 4 \epsilon} { 1 \over
\sqrt{c_{1}^3 c_{2}}}
\Gamma_{\mu \nu}^{2}(x) \, ,
\eea
and since
$\Gamma_{\mu \nu}^{2} + \Delta \Gamma_{\mu \nu}^{2}
\equiv \Gamma_{\mu \nu}^{(r) 2} = Z_{B}^{-1} \Gamma_{\mu \nu}^{2}$,
this implies
\bea
Z_{B}^{-1} &=& 1+ {\bar{\lambda}^{2}  \over 4 \epsilon} { 1 \over
\sqrt{c_{1}^3 c_{2}}}  \,\, . \label{zb}
\eea
As $ \lambda = \lambda_{r} + O(\lambda_{r}^{3})$, therefore
\bea
Z_{B} &=& 1 - {\bar{\lambda}^{2}_{r}  \over 4 \epsilon} { 1 \over
\sqrt{c_{1}^3 c_{2}}}  \,\, . \label{zb}
\eea

To motivate the choice (\ref{c1}-\ref{c2}), let us assume that
$h_{n}$ and $f_n$, for large $n$, have the power-like
asymptotic forms
\bea
h_n &\to & c_{s} |n|^{s}  \label{s1} \, ,\\
f_n &\to & c_{t} |n|^{t}  \label{s2} \, .
\eea
For integral $s,t > 0$, the sum (\ref{sum}) for large $n$ is then
\bea
&\sim& \sum_{n} |n|^{d-2} \ |n|^{t({d-4 \over 2})} \ |n|^{-sd
\over 2} \, .
\eea
{}From the properties of the Riemann $\zeta$-function,
as $d \to 3$, this has a divergence only if
\bea
3s+t =4 \; . \label{gen}
\eea
The only positive integral solution to (\ref{gen}) is $s=t=1$.
If the condition
(\ref{gen}) is not satisfied then the $n$ sum is finite as $d \to 3$
and we would obtain $Z_B =1$.
Furthermore as $Z_5 =1$ (see next subsection) independent
of $f_n$ and $h_n$, this implies that the one-loop $\beta$ function
for $\lambda$ is zero for the { \it ansatz} (\ref{s1}-\ref{s2}) unless
$s=t=1$. Thus our choice (\ref{c1}-\ref{c2}), and any others equivalent
to it at large $n$, would give a running coupling while other choices
of $h_n$ and $f_n$ with asymptotic forms (\ref{s1}-\ref{s2}) would
give a UV finite theory at one-loop.

\subsection{$Z_{5}$}
This is determined from the two diagrams contributing to the
self-energy of the $A_5$ field :
The tadpole diagram Fig.(3a)
and the non-tadpole Fig.(3b)
diagrams. Neither of these involves an internal $n$ summation. Since the
$a_\mu$ propagator is massive, the tadpole digram has a potential UV divergence
but this divergence is linear and vanishes in dimensional regularization:
Recall that in dimensional regularization (DR) the  scaleless integral
($d >2$)
\bea
\int {d^d q \over q^2} &=& 0 \, ,
\eea
so that
\bea
\int {d^d q \over q^2 +m^2}&=& \int d^d q \left( {1 \over q^2 + m^2}
- {1 \over q^2} \right) \nonumber \\
&=& \int {d^d q \ (-m^2) \over q^2 (q^2+m^2) }
\eea
is finite as $d \to 3$.
Similarly a  potential linear UV divergence in the other diagram is
removed in  DR.  (We mention in passing that as $p \to 0$, the
UV finite pieces of the two diagrams cancel so that the
$A_5$ field self-energy vanishes at zero momentum, as required by  gauge
invariance.) Hence $Z_5=1$ at one-loop order. As a cross-check,  we
have  verified, by looking at the
$<A_{\mu} A_{\nu}>$ and $<A_{\mu} A_5>$
self-energies that they are also finite near $d=3$ so that $Z_5 =Z_{A_\mu}
=1$ as required by gauge-invariance in the background gauge.
Furthermore, the fact that the divergence in the
$B_\mu$ field two-point funstion could
be renormalised by a wavefunction renormalisation and that the
$A_\mu$ and $A_5$ two-point functions are finite,
justifies {\it a posteriori} our  assumption in the
last subsection
of multiplicative renormalizibility of the two-point functions at one-loop.
Thus it follows from (\ref{lo}) and (\ref{zb})
that
\bea
\beta_{\lambda} &=& { - \bar{\lambda}_{r}^{3}
\over 4 \sqrt{c_{1}^{3} c_{2} }}
\; .
\eea
The pure, abelian, Kac-Moody gauge theory is asymptotically free !
The
only other known asymptotically free gauge theory in four dimensions
is Yang-Mills (YM) theory (and linearizations thereof \cite{N}),
which is based on a non-abelian group.
It has been understood \cite{N} that the crucial ingredient
for  asymptotic freedom in YM theory is the self-interaction of spin
one (vector)
fields which makes the vacuum of YM theory behave like a paramagnetic
medium.  As the { \it abelian}
Kac-Moody gauge theory also has  self-interacting
vector fields, asymptotic freedom in this case has probably a similar
physical interpretation. However we remind that unlike YM theory,
the Kac-Moody gauge theory is  nonlocal, or equivalently
it may be interpreted as a local theory but with an infinite set
of non-commuting charges.

\subsection{Including Dirac Fermions.}
The Dirac fermions $\psi(x, \sigma), \bar{\psi}(x, \sigma)$
couple only to the
gauge-fields $A_{\mu}, A_5$ as indicated in (\ref{1f}), and
the resulting theory with gauge-fields and Dirac fermions
has two independent
coupling constants : $\lambda$ and $e$.
The wave-function renormalization constant $Z_A$ now obtains a
quantum-electrodynamic-like contribution \cite{R}
from the fermion-loop
diagram of Fig.(4),
\bea
Z_5=Z_A=1- {e^2 N \over 12 \pi^2 \epsilon} \; , \label{newz}
\eea
where $N$ is the number of fermion flavours. The divergence which
(\ref{newz}) absorbs (through a  rescaling of the bare fields $A_\mu, A_5$)
is that in the loop correction to the usual kinetic terms
${1 \over 4} F_{\mu \nu}^2$ and
${1 \over 2} F_{\mu 5}^2$, that is, Eq.(\ref{class}) with
$h(\sigma - \sigma^{\prime})=f(\sigma-\sigma^{\prime})
= \delta(\sigma-\sigma^{\prime})$. Therefore with Dirac fermions present,
the renormalizability of the theory seems to restrict the choice
(\ref{c1}-\ref{c2}) further to $h_n=f_n =1$. However, as we argue below,
 it is still possible to
choose the  more general form
\bea
h_n = f_n = 1+ c|n| \label{mg} \, .
\eea
With (\ref{mg}),
the kinetic terms for the $A_{\mu}$ and $A_5$ fields in (\ref{class})
are now split into two pieces each : The $'1'$ in (\ref{mg}), gives rise
to the conventional kinetic term which is renormalized by (\ref{newz})
while the $c|n|$ part gives rise to another kinetic term. Since at one-loop
there is no ultraviolet divergence corresponding to the latter
kinetic term, consistency demands that $c$ may be
nonzero only if it renormalizes inversely to (\ref{newz}), so that
the unconventional kinetic term is unrenormalized. That is, we require
\bea
Z_c Z_{A} &=& 1 \, , \label{zca}
\eea
where the bare parameter $c$ is related to the renormalized parameter
$c_r$  by
\bea
c &=& Z_c c_r \label{zc} \, .
\eea
(Actually what is being renormalized is a mass parameter $c/R$).

Now the bare charge $e$ is related to the renormalized charge $e_r$
by
\bea
e &=& \left( {\mu \over \mu_{0}} \right)^{\epsilon} Z_{e} e_r \label{ze} \, .
\eea
Since from (\ref{1f}) and gauge-invariance in the background gauge
\bea
Z_e &=& {1 \over \sqrt{Z_{A}}} = {1 \over \sqrt{Z_{5}}} \, , \label{ze2}
\eea
therefore \cite{R}
\bea
\beta_{e} &=&
\mu {\partial e_r \over \partial \mu} \nonumber \\
&=& -e_r \epsilon \ + \ { e_{r}^{3} N \over 12 \pi^2} \, ,
\eea
as in quantum electrodynamics (QED). (We have kept the term of order
$\epsilon$ in $\beta_{e}$ as it is needed below). Then one obtains
\bea
\gamma_c &\equiv &
\mu {\partial c_r \over \partial \mu} \nonumber \\
&=&  { c_{r} e_{r}^{2} N \over 6 \pi^2} \, .
\eea
Note that there is no term of order $\epsilon$ on the right-hand-side of
the last equation because, as mentioned before,
$c/R$ is a mass parameter of fixed mass dimension 1, so that no
compensating mass parameter $\mu^{\epsilon}$ enters in its dimensional
continuation.

Using the values of $\beta_e, \gamma_c, Z_5$ above,
 and $Z_b$ from (\ref{zb})
with $c_1=c_2=c$, we obtain from (\ref{lo})
\bea
\beta_{\lambda} &=&  - { \bar{\lambda}_{r}^{3} \over 4 c_{r}^{2}} \ + \
  { \bar{\lambda}_{r} e_{r}^{2} N \over 6 \pi^2} \,
\label{bnew}
\eea
to lowest order. The contribution of the fermions is of opposite sign to that
of the gauge-fields just as in quantum-chromodynamics (QCD). However unlike
QCD, we have two independent
coupling constants appearing in (\ref{bnew}) so that
$\beta_{\lambda}$ can be arranged to have a nontrivial zero in weak coupling
: For $e_r$ small, $\beta_{\lambda}$ vanishes at
\bea
\bar{\lambda}_{r}^{\star} =
e_r \ {2c_r \over \pi} \left({N \over 6} \right)^{1 \over 2}
\, .
\label{ll}
\eea
Since  $\bar{\lambda}_{r}^{\star} \sim O(e_r)$
even for  $c_r$ of order 1 and $N$ not too large,
the result (\ref{ll}) is self-consistent
as it is within the perturbative regime for
{\it both} $\bar{\lambda}_r$
and $e_r$. Thus starting from short distances (but $e_r$ still small),
$\bar{\lambda}_r$ tends to increase at longer distances, but before it
can become too large the fermion coupling drives it to
$\bar{\lambda}_{r}^{\star}$. This point is infrared stable since
$\beta_{\lambda}^{\prime}(\bar{\lambda}_{r}^{\star}) > 0.$
However $\bar{\lambda}_{r}^{\star}$
is not a fixed point of the whole system since $e_r$ continues to run.

\setcounter{equation}{0}
\section{Renormalizability}
We want to show that the theory as defined by (\ref{qact})
is at least one-loop
renormalizable. That is, for all  one-loop diagrams with
an arbitrary number of external legs, the UV divergences
can be absorbed by rescaling  the terms in the lagrangian
(\ref{class}). Note that
since the
tensors $\Gamma_{\mu \nu} , F_{\mu \nu}, F_{\mu 5}$ are individually
gauge-invariant, radiative corrections could in principle
generate more gauge-invariant
terms than are in  the lagrangian (\ref{class}), such as
\bea
\Delta {\cal{L}} &=& \int \left\{ \alpha_{1}
(\partial_{\mu} F_{\mu \nu}) F_{\nu 5} \ + \ \alpha_{2} \Gamma_{\mu \nu}
F_{\mu \nu} \ + \ \alpha_{3} F_{\mu \nu} \partial^2 F_{\mu \nu} \ + \
......\right\}
\label{nct}
\eea

In this section we show that, remarkably, all the one-loop UV divergences
have the same form as the minimal lagrangian (\ref{class})
so that no extra terms such
as (\ref{nct}) are needed for renormalization.
This is also fortunate because some of
the terms in (\ref{nct}) are not positive definite and so unsuitable for
inclusion in the tree-level lagrangian. More specifically, we will
sketch renormalizability of the theory for the
case
\bea
h_n &=& c_{1} |n| + c_{1}^{\prime} \\
f_n &=& c_{2} |n| + c_{2}^{\prime}
\eea
discussed in the last section (for (\ref{s1}-\ref{s2})
the theory is even more convergent).
For the purpose of determining the UV degree of divergence of the one
loop diagrams, it is sufficient to consider the large $n$ limit in
which case we can treat the discrete sum as a continuous integral by
making the replacements
\bea
 {n \over R} \to q_z \;\;\; \mbox{and} \;\;\;  {1 \over R} \sum_{n} \int d^3 q
\to \int dq_z d^3 q \; . \label{eff}
\eea
In sections 5.1 and 5.2  we consider
various subsets of one-loop diagrams contributing to
the 1PI action to the pure gauge theory and in section 5.3 we
consider the fermions (\ref{1f}).

\subsection{Case A: diagrams containing at least one external $\Gamma$ vertex}
(i) Exactly two $\Gamma$'s :\\
We have already shown (in Sect 4.2) that the one-loop diagrams with
two $\Gamma (\equiv \Gamma_{\mu \nu})$ vertices has a logarithmic
UV divergence which is removed by wavefunction renormalization, and that
no other divergences remain in those diagrams.

(ii)More than two $\Gamma$'s:\\
Since the two-$\Gamma$ diagrams were logarithmically divergent, adding any
more external  vertices makes the diagrams UV convergent.

(iii) One $\Gamma$ and some other external vertices:\\
{}From the Feynman rules, one sees that at one-loop
no gauge-invariant term in the 1PI action can be formed
from exactly one $\Gamma$ and one $F_{\mu \nu}$ (or $F_{\mu 5}$). So consider
possible terms of the form $\Gamma (F)^{n}, n \ge 2$, (we have
used a symbolic notation : indices and possible derivatives
have been suppresed,  and $F$ here denotes $F_{\mu \nu}$ or $F_{\mu 5}$).
Consider first $n=2$; Some
of these diagrams (see Fig.(5))
actually contribute to $\Delta \Gamma^{2}_{\mu \nu}$
but by gauge-invariance their UV divergence is
related to the wavefunction renormalization of the
$B_\mu$ field already discussed in  case A(i) above.
There are also diagrams which are not part of
corrections to $\Gamma_{\mu \nu}^{2}$
and so must be part of corrections to $\Gamma (F)^2$ (see e.g. Fig.(6)).
We wish to show that
the net contribution of diagrams such as in Fig.(6) to $\Gamma (F)^2$
is only a finite renormalization. Note that superficially Fig.(6) is
logarithmically UV divergent but gauge-invariance demands that the external
$A_5$ vertex must eventually (either after doing the detailed calculation
or after adding other similar diagrams) become a $\partial_{\mu}A_5$.
This then makes the diagram  UV convergent.

With the above effective power counting (that is,
(\ref{eff}) plus gauge invariance), one can show that all
diagrams
contributing to $\Gamma (F)^2$ are convergent. Then it follows
that $\Gamma (F)^n \, (n \ge 2)$, radiative contributions
to the 1PI action at one-loop order
are UV finite.

\subsection{Case B: diagrams {\it not}
containing an  external $\Gamma$ vertex.}
(i) Terms of the form $(F)^{2n+1}, n \ge 1$.\\
It is impossible to form such terms at one-loop order as it requires an odd
number of $A-b-a$ vertices which cannot be contracted since the $b$ fields
must occur in pairs.

(ii) $(F)^4$\\
Some of the diagrams containing four external $A$ lines (see eg.
Fig(7)) contribute
to $\Delta \Gamma_{\mu \nu}^{2}$ and  therefore their UV divergence
is as before related to the wavefunction renormalization of the
$B_\mu$ field. Here we need to consider those four-$A$ diagrams
which
contribute to $F^4$. Consider for
example the diagram of Fig.(8). There is no internal $n$-sum because of
the $b$ propagator, so superficially the UV degree of divergence is
that of the integral $\int {d^3 q \ q^3 \over q^6}$ which
is logarithmic.
More explicitly integrals of the form
\bea
\int {d^d q \ q_{\mu} q_{\sigma} q_{\alpha} \over q^4 (m^2 + (p-q)^2)}
\eea
appear but these are finite as $d \to 3$ because in an expansion around
$p=0$, the first term is zero by symmetry and the rest are finite.
Thus the superficially divergent diagram is actually UV finite.

(iii) $F^{2n}, n \ge 3$. \\
Since the $n=2$ diagrams were only (superficially) logarithmically
divergent, the  $n \ge 3$ diagrams are therefore
 convergent by the  effective power counting
rules discussed above.

\subsection{Diagrams with Dirac fermion lines}
Since the fermionic part of the action (\ref{1f}) looks like QED,
all the
one-loop diagrams are the same as there. As in QED, there is only one UV
divergent diagram with a closed fermion loop, Fig(4). However unlike
QED, the fermion self-energy diagram and the vertex diagram in  Fig.(9)
are UV finite  for $d=3$ because the $a_\mu$ propagator is
more damped (see eq.(\ref{aprop})) than the photon in normal QED.  Thus the
vertex renormalization factor $Z_v$ and the fermion wave-function
renormalization factor $Z_{\psi}$ corresponding to Fig.(9) equal identity.

\setcounter{equation}{0}
\section{Conclusion}
We have studied in this paper the quantization of
Kac-Moody gauge-field theory introduced in
Ref.\cite{B1}. Although such a theory had been written in \cite{B1}
for the non-abelian case, we considered here the abelian limit which
still retained the interesting features of nonlinerality, nonlocality
and spatial asymmetry. The last two features are a result of
interpreting the
internal Kac-Moody space as a compact spatial coordinate.

We quantized the pure $\hat{U}(1)$ Kac-Moody gauge
theory (\ref{class}) using the path-integral formulation
in a background field and showed that it was one-loop renormalizable.
Indeed, at one-loop, we found that because of the unusual propagators
for the $a_\mu$ and $a_5$ fields, the theory was UV finite
within dimensional
regularisation for a wide range of the functions $h_n$ and $f_n$
appearing in the classical theory. However for $h_n$ and $f_n$ behaving
like $|n|$ for large $n$, we found the
dimensionless coupling $\bar{\lambda}_r$ was asymptotically free. When Dirac
fermions (\ref{1f}) were added, the theory gained another coupling constant,
$e$. At first sight,
renormalizability and gauge-invariance now seemed to  demand that the
functions $h_n$ and $f_n$ took the form  (\ref{mg}) with $c=0$. However
by keeping $c$ nonzero, we obtained a more interesting
structure for the beta-functions: This could be done by promoting
$c$ from a fixed quantity to a
mass-like parameter which was allowed to be renormalized
according to (\ref{zca}). In this way the constraints of gauge-invariance
and renormalizability could be satisfied and the beta function
$\beta_{\lambda}$  obtained two terms of opposing sign (\ref{bnew}).
The couplings $e_r$ and $\bar{\lambda}_r$ could be chosen in the
perturbative regime to make $\beta_{\lambda}$  vanish.

If the theory is to be regarded as a fundamental theory then its
renormalizability should in the future be studied to all orders.
It would also be interesting to quantize the theory by including the
Kac-Moody fermions (\ref{2f}). We only note here
from the actions (\ref{class}),
(\ref{1f}) and (\ref{2f}) that the partition function for the full theory
comprising gauge-fields and Kac-Moody fermions has the intriguing property of
apparently being sensitive to the sign of the parameter $\kappa$ appearing in
the Kac-Moody algebra (\ref{al}).

One might wonder about the theory for other values of $d$. We found that
already at one-loop the theory (\ref{qact}) was not renormalizable for
$d=4$ : For example a divergence of the form $p^4 R^2 / (d-4)$ appeared
in the self-energy of the  $A_5$ field and such a term could not be
absorbed by wave-function renormalization. For $d=1$ and $d=2 $
 the pure
gauge-theory is free.

Even if the $3+1$ dimensional theory is not ultimately renormalizable
to all orders in the sense of QED, it might be profitable to treat it
as a low energy effective theory.
As the lagrangian (\ref{class}) is naturally
asymmetric, one is tempted then
to associate it with a description of
 high-temperature superconductors (HTS),
which are anisotropic systems.\footnote{At non-zero temperature
the Kac-Moody gauge theory would be defined on ${\cal{S}}^1 \times
{\cal{R}}_{d-1} \times {\cal{S}}^1$.} It has been
argued (see \cite{P} and references therein})
that the non-Fermi liquid behaviour of the normal state of HTS requires for
its explanation a planar dynamical gauge field in the effective low energy
theory. This role  in the lagrangian (\ref{class}) could be played
by $B_{\mu}(x)$, while $A_{\mu}(x, \sigma), A_{5}(x,\sigma)$
and the fermions would describe some other fields (perhaps the
effective electromagnetic fields and fermionic quasi-particles respectively).
Of course detailed calculations are required to see if our proposal of
(\ref{class}) has indeed the required properties to describe the
properties of HTS in the normal state, and if it has a phase transition
at non-zero temperature to
a superconducting state. In any case, regardless of a particular
application, it would be interesting to study how the infrared dynamics
of (\ref{class}) depends on the anisotropy scale $R$, and what physical
consequences issue from the Kac-Moody symmetry of the system.\\ \\

\noindent
{\bf Acknowledgement}\\
R.P. thanks the High Energy Theory Group at the National University of
Singapore for hospitality during the course of this work.\\

\vspace*{1cm}


\newpage

\vspace*{0.5cm}
{\bf Figure Captions}\\

Fig.1. The $a_\mu$ propagator (wavy line),
the $a_5$ propagator (curly line), the $b_\mu$ propagator (dashed
line) and the Dirac fermion propagator (unbroken line).\\

Fig.2. Diagrams contributing to $Z_B$.\\

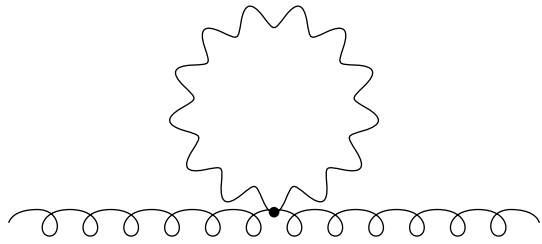
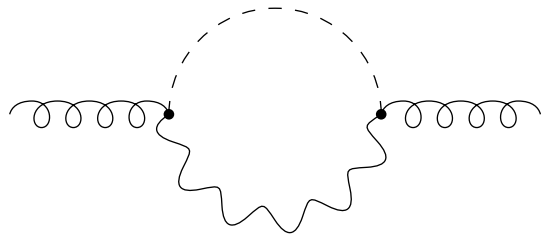
Fig.3. Gauge field diagrams contributing to $Z_5$.\\

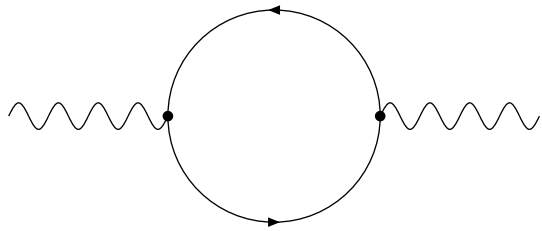
Fig.4. Fermion loop contributing to $Z_5$.\\

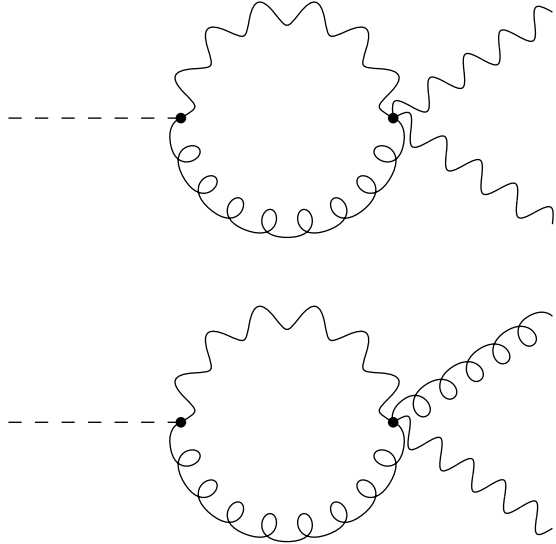
Fig.5. UV divergent
diagrams contained in the one-loop  correction to $\Gamma_{\mu \nu}^{2}$.\\

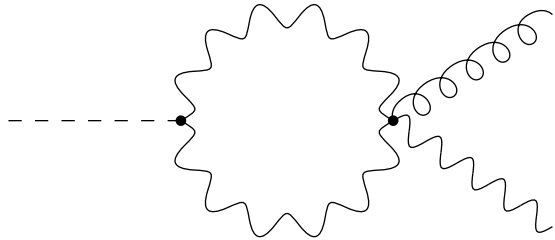
Fig.6. A diagram not contained in $\Gamma_{\mu
\nu}^2$. The curly external line is $A_5$, the wavy external line is
$A^{\prime}_{\mu}$, while the dashed external line is $\partial_{\mu}
B_{\tau}$.\\

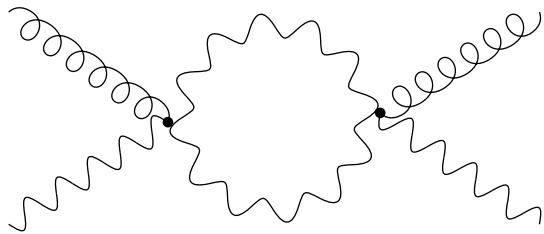
Fig.7. A diagram contained in the one-loop
correction to $\Gamma_{\mu \nu}^{2}$.\\

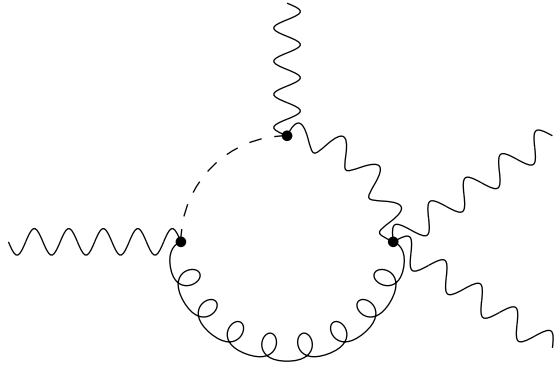
Fig.8. A superficially UV divergent
diagram not contained in the one-loop correction to $\Gamma_{\mu
\nu}^2$. It is UV finite.\\

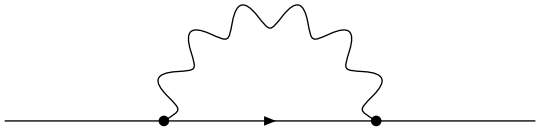
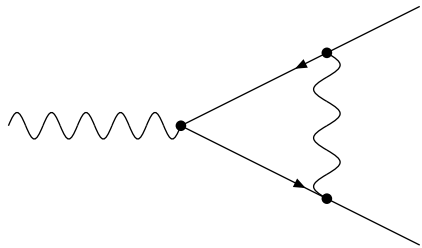
Fig.9. Fermion self-energy
and vertex-correction diagrams.\\


\begin{thebibliography}{99}
\bibitem{PG} P. Goddard and D. Olive, "Kac-Moody and Virasoro
Algebras",\\ (World Scientific, 1988).

\bibitem{B1} B.E. Baaquie, Phy. Lett B271 (1991) 343.

\bibitem{B2} B.E. Baaquie, hep-th/9511094, to appear in Phys. Rev. D
(1995).

\bibitem{HA} G. 't Hooft, Nuc. Phys. B62 (1973) 444; \\
L.F. Abbott, Nuc. Phys. B185 (1981) 189.

\bibitem{N} N.K. Nielsen, Am. J. Phys. 49 (1981) 1171.

\bibitem{R} L.H. Ryder, "Quantum Field Theory", \\
(Cambridge University Press, 1994).

\bibitem{P} J. Polchinski, Nuc. Phys. B422 (1994) 617.


\end{thebibliography}
\end{document}